\documentstyle[preprint,aps,epsf]{revtex}
\newcommand{\vev}[1]{\langle #1 \rangle}
\newcommand{\vd}{|(V_{tb} V^*_{td})/(V_{ub} V^*_{ud})|}
\newcommand{\vs}{|(V_{tb} V^*_{ts})/(V_{ub} V^*_{us})|}
\begin{document}
\draft
%\twocolumn[
%\preprint{\vbox{
%\null\hfill KEK preprint 95-xxx \\
%\hfill KEK-TH 476\\
%\vskip -2cm\noindent %\epsfbox{kekm.eps}
%}}
\preprint{\vbox{\hbox{KEK preprint 96-33}\hbox{KEK TH-476}
\hbox{FERMILAB-PUB-96/130-T}}}
\title{
\vskip -3cm\epsfbox{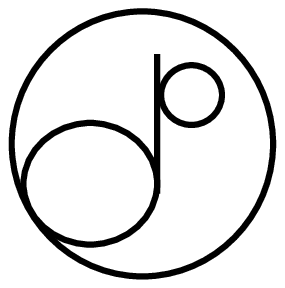}
  Direct $CP$ violations of $B$ meson via $\rho-\omega$ interference
\footnote{submitted for publication.}
}
\author{
  Ryoji Enomoto$^a$\footnote{e-mail: {\tt enomoto@kekvax.kek.jp.}}
and
  Masaharu Tanabashi$^{a,b}$\footnote{e-mail: {\tt tanabash@theory.kek.jp}}
}
\address{
$^a$National Laboratory for High Energy Physics, 
1-1 Oho, Tsukuba, Ibaraki 305, Japan\\
$^b$Fermilab, MS 106,
P.O.Box 500, Batavia, IL 60510\footnote{Address till August 25, 1996.}
}

\date{\today}
\maketitle
%\widetext
\begin{abstract}

We investigate $B^{\pm,0}\rightarrow \rho^0 (\omega )h^{\pm,0}$, where  
$\rho^0 (\omega)$ decays to $\pi^+\pi^-$ and $h$ is
any hadronic final states, such as $\pi$ or $K$.
We find large direct $CP$ asymmetries via $\rho-\omega$ interference.
A possible determination of the weak phases, such as  
$\phi_2={\rm arg}((V_{ud}V^*_{ub})/(V_{td}V^*_{tb}))$ 
and
$\phi_3={\rm arg}((V_{us}V^*_{ub})/(V_{ts}V^*_{tb}))$,
is also
discussed. 
We show the feasibility to detect the CP asymmetries in these channels
by assuming an asymmetric $e^+e^-$ collider experiment.
$10^{9}$ $B\bar B$ events are required for the detection of this effect.

%Feasibility of detecting this effect is also discussed.
%Any B-factories such as $e^+e^-$ (symmetric or asymmetric collider) and
%hadron machine can detect its asymmetry if they have smaller mass
%resolution of $\pi^+\pi^-$ than the width of $\omega$ (8.4 MeV)
%at around $\omega$ mass and large statistics such as $10^8B\bar{B}$.
\end{abstract}
%]
%\newpage
%\narrowtext

%introduction
Although $CP$ violation has been observed in the $K^0$-$\bar K^0$
system since 1964, we have no evidence 
up to now of $CP$ violation other than
in the kaon system.
For understanding the origin of the $CP$ violation, however, we
definitely need information concerning $CP$ violations other than 
that in the
$K^0$-$\bar K^0$ system.

The standard model of the $CP$ violation\cite{km} predicts large $CP$
asymmetries in the $B$ meson system\cite{sanda}.
Many experimental attempts to detect $CP$ violations of the $B$ meson will
be carried out towards the next century\cite{belle,babar,herab}.
Thanks to the large number of its decay modes, a precise
measurements of the $B$-meson $CP$ asymmetries will supply much
information for a deeper understanding of the origin of the $CP$
violation, which will provide clues for the new physics beyond the
standard model.
For such a purpose, we need to consider the various $B$-meson $CP$
asymmetries (as many as possible).

Unlike the $CP$ violations in the neutral $B$ meson, the $CP$
asymmetry of the charged $B$ meson can be caused solely by
direct $CP$ violations, which only occur through 
interference between two amplitudes having different weak and strong
phases. 
In the standard model a weak-phase difference is provided by the
different complex phase of the Kobayashi-Maskawa (KM) matrix
elements\cite{km} of the tree and penguin diagrams, while the strong phase
is given by the absorptive parts of the corresponding diagrams. 

To obtain a large direct $CP$ asymmetry, which can be observed in future
$B$ factories, we need to consider a decay mode with a sufficiently 
large strong phase difference.
We also need to know precisely the size of the strong-phase difference
in order to 
extract the size of the weak phase difference (KM phase) from the
observed direct $CP$ asymmetry.

However, it is difficult to control such a large absorptive part (strong
phase) in a perturbative manner.
Nonperturbative resonance states, in which we know the behavior of 
the absorptive part by using the Breit-Wigner shape, are ideal places for
this game.
The $CP$ violations via the radiative decays of $B$ meson were predicted
by Atwood and Soni\cite{gammak,gammaa}.
The role of charmonium resonances in the $CP$ violation of $B^\pm$ decay
was discussed by Eilam, Gronau and Mendel\cite{eilam}. 
%They found a 10\% level $CP$ asymmetry in the resonance background
%interference. 
Lipkin discussed the use of $\rho$-$\omega$ interference as a trigger
of the direct $CP$ violation in neutral $B$-meson decay using
a simple quark model analysis \cite{kn:lipkin}.

In this Letter we present a systematic analysis of the $CP$ asymmetries in
$B^{\pm, 0}\rightarrow \rho^0 (\omega) h^{\pm,0} \rightarrow
\pi^+\pi^- h^{\pm,0}$  
via $\rho \omega$ interference\cite{nambu}, 
where $h$ is any hadronic final state, such as $\pi$, $\rho$, $K$,
$K^*$.
We find large $CP$ asymmetries at the interference region.
We also show the feasibility of this method assuming a realistic luminosity
and detector.
%Two diagrams having different KM phases $\phi$ contribute to this
%decay mode (see Figures \ref{feynman} (a) and (b)). 
%The intermediate $\omega$ 
%state gives a large $\delta$ ($\sim$90 degrees).
Although the size of $\rho$-$\omega$ mixing is quite small, 
since it is triggered by the small isospin violation, 
an enhancement at the $\omega$ pole can overcome this smallness
at the $\omega$ pole region. 
%thanks to the narrow width of $\omega$ meson.  
%Penguin amplitude ($\Gamma _{\rho}m_{\omega}/\Gamma _{\omega}^2$)
%with respect to the direct decay of $\rho^0$ in the tree diagram. 
In addition, these decay modes are considered to have a 
reasonably large branching fraction, such as $10^{-5}$\cite{br1,br2}.

\begin{figure}
\epsfbox{feynman.eps}
\vskip 1cm
\caption{Examples of the Feynman diagrams of the decay
$B^-\rightarrow \rho^0 (\omega) \rho^-$;
(a) tree and (b) penguin diagram.}
\label{feynman}
%\newpage
\epsfysize 16cm
\epsfbox{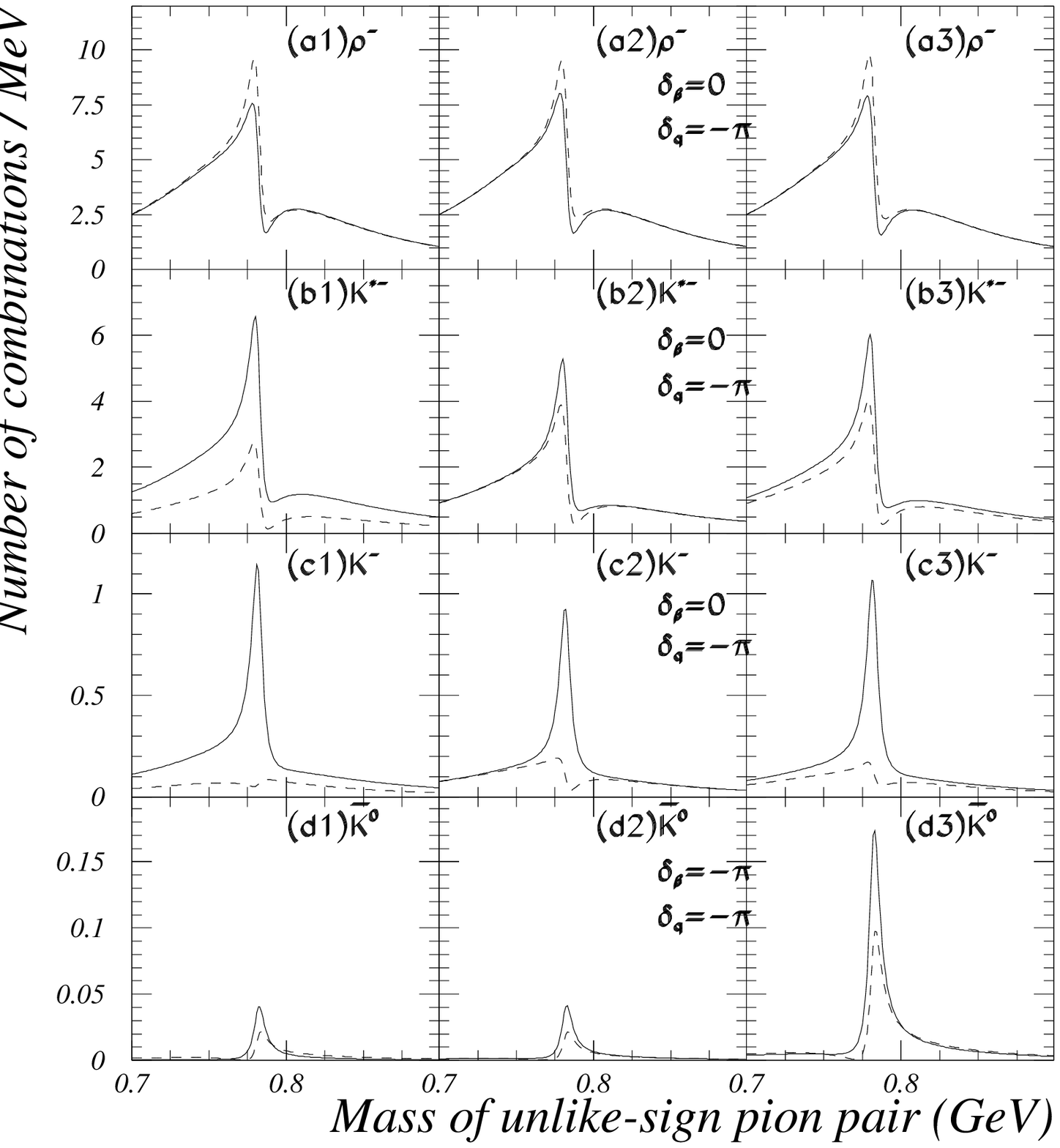}
\caption{
Expected invariant mass spectra of unlike-sign pion pairs.
The solid lines are for $B^+$ or $B^0$ decays,
and the dashed ones are for $B^-$ or $\bar{B^0}$
decays. 
The vertical scale is the differential yield for  
$(\pi^+\pi^-)h^{\pm ,0}$
combinations, and is normalized to give 
the number of entries at $10^8~B\bar{B}$ events, assuming an
100-\% acceptance.
The details concerning the notations (a1)-(d3) are described in the text.
}
\label{pattern}
\end{figure}

%simple case
Figures \ref{feynman} (a) and (b) are examples of quark level diagrams
of the tree (a) and penguin (b) amplitudes for 
$B^{\pm}\rightarrow \rho^0 (\omega) \rho^{\pm}$.
Considering the quark components: while diagram (a) gives a final
state of $\rho^0+\omega$, diagram (b) contributes solely to
a $\omega$ meson. 
The standard model predicts a weak phase difference, 
$\phi_2 = {\rm arg}((V_{ud}V^*_{ub})/(V_{td}V^*_{tb}))$.
The absorptive part (strong phase) is provided by both the
$\rho$-$\omega$ interference and the Bander-Silverman-Soni
mechanism (the quark loop absorptive part in the penguin
diagram)\cite{kn:BSS}. 

We start with a general strategy for evaluating the $CP$ asymmetry
in this system. The $CP$ asymmetry is given by
\begin{equation}
  a \equiv \frac{|A|^2 - |\bar A|^2}{|A|^2 + |\bar A|^2}
    = \frac{ r (\cos(\delta + \phi) - \cos(\delta-\phi))}
           {1 + r^2 + r (\cos(\delta+\phi)+\cos(\delta-\phi))},
\label{eq:cpasym}
\end{equation}
where $\delta$ and $\phi$ are the strong and weak phases, respectively.
The amplitudes ($A$ and $\bar A$) are
\begin{eqnarray*}
  A      &\equiv& \vev{\pi^+ \pi^- h | {\cal H}^T | B}
                 +\vev{\pi^+ \pi^- h | {\cal H}^P | B}
          = \vev{\pi^+ \pi^- h | {\cal H}^T | B}\left[
                1 + r e^{i\delta} e^{i\phi} \right],
  \\
  \bar A &\equiv& \vev{\pi^+ \pi^- \bar h | {\cal H}^T | \bar B}
                 +\vev{\pi^+ \pi^- \bar h | {\cal H}^P | \bar B}
          =  \vev{\pi^+ \pi^- \bar h | {\cal H}^T | \bar B}\left[
                1 + r e^{i\delta} e^{-i\phi} \right],
\end{eqnarray*}
with ${\cal H}^T$ and ${\cal H}^P$ being the tree and penguin Hamiltonians,
respectively. 
The parameter $r$ stands for the absolute value of penguin/tree ratio, 
\begin{equation}
  r \equiv \left| \frac{\vev{\pi^+ \pi^- h | {\cal H}^P | B}}
                       {\vev{\pi^+ \pi^- h | {\cal H}^T | B}}
           \right|.
\end{equation}

We assume that the vector meson contributions dominate over the
continuum $\pi^+ \pi^-$ final state when the $\pi^+ \pi^-$ invariant
mass  is sufficiently close to the $\omega$ meson mass
$m_\omega$, 
\begin{equation}
  \vev{\pi^+\pi^- h | {\cal H}^{T,P} | B} 
  \simeq \vev{\pi^+\pi^-|J_0^\mu|0}\frac{\epsilon_\mu(\omega)}{g_\omega}
         \vev{\omega h|{\cal H}^{T,P} | B}
        +\vev{\pi^+\pi^-|J_3^\mu|0}\frac{\epsilon_\mu(\rho)}{g_\rho}
         \vev{\rho h|{\cal H}^{T,P} | B},
\end{equation}
where $\epsilon_\mu(\omega/\rho)$ and $g_{\omega/\rho}$ are the
polarization vector and the decay constant of $\omega/\rho$,
respectively. 
This assumption can be confirmed experimentally.
For later convenience, we define $\alpha$, $\delta_\alpha$,
$\beta$, $\delta_\beta$, $r'$ and $\delta_q$ as:
\begin{equation}
\alpha e^{i\delta_\alpha}
  \equiv 
         \frac{\vev{\omega h|{\cal H}^{T} | B}}
                     {\vev{\rho h|{\cal H}^{T} | B}},
         \qquad
\beta  e^{i\delta_\beta}
  \equiv 
         \frac{\vev{\rho h|{\cal H}^{P} | B}}
                     {\vev{\omega h|{\cal H}^{P} | B}}, 
         \qquad
  r' e^{i(\delta_q+\phi)}
         \equiv 
         \frac{\vev{\omega h|{\cal H}^{P} | B}}
                    {\vev{\rho h|{\cal H}^{T} | B}},
\end{equation}
with $\delta_\alpha$, $\delta_\beta$ and $\delta_q$ being the strong
phases (absorptive part) at short distance.
These strong phases can be roughly evaluated from the quark loop
diagrams\cite{kn:BSS}.
%We expect existence of much larger absorptive part in the 
%hadronic $\pi\pi$ final state interaction in 
%$\vev{\pi^+\pi^-|J_{0,3}^\mu|0}$.  
%Hereafter we neglect small imaginary parts of $\alpha$ and $\beta$ for
%simplicity. 
By using these parameters, we obtain
\begin{equation}
  r e^{i\delta} =r' e^{i\delta_q} 
                  \frac{\beta e^{i\delta_\beta} 
                              \vev{\pi^+ \pi^-|J_3^\mu|0}
                              \frac{\epsilon_\mu}{g_\rho}
                             +\vev{\pi^+ \pi^-|J_0^\mu|0}
                              \frac{\epsilon_\mu}{g_\omega}}
                       {      \vev{\pi^+ \pi^-|J_3^\mu|0}
                              \frac{\epsilon_\mu}{g_\rho}
                       +\alpha e^{i\delta_\alpha}
                              \vev{\pi^+ \pi^-|J_0^\mu|0}
                              \frac{\epsilon_\mu}{g_\omega}}.
\label{eq:stphase}
\end{equation}

We now evaluate the long-distance hadronic $\pi^+\pi^-$ final
state interaction  
at the $\omega$ meson mass region using a simple model of
$\rho$-$\omega$ mixing\cite{qmass}: 
\begin{eqnarray*}
\vev{\pi^+\pi^-|J_0^\mu|0} 
  &\sim& g_{\rho\pi\pi} \frac{1}{m_\rho^2-i\Gamma_\rho m_\rho - s}
          \left[ \frac{g_\omega}{3} \frac{e^2}{-s} g_\rho+g_{\rho\omega}
          \right]
          \frac{1}{m_\omega^2-i\Gamma_\omega m_\omega - s} g_\omega, 
  \\
\vev{\pi^+\pi^-|J_3^\mu|0} 
  &\sim& g_{\rho\pi\pi} \frac{1}{m_\rho^2-i\Gamma_\rho m_\rho - s}
  g_\rho,
\end{eqnarray*}
where $|\omega\rangle$ and $|\rho\rangle$ are defined as the isospin
eigenstates and $g_\omega$, $g_\rho$, $g_{\rho\omega}$ and
$g_{\rho\pi\pi}$ are the decay constants of the $\omega$ and $\rho$ mesons,
$\rho$-$\omega$ mixing amplitude and $\rho\pi\pi$ coupling,
respectively.
We thus obtain
\begin{equation}
\frac{\vev{\pi^+ \pi^-|J_0^\mu|0}\epsilon_\mu}
     {\vev{\pi^+ \pi^-|J_3^\mu|0}\epsilon_\mu}
  \simeq
% \frac{e^2}{3} \frac{g_\omega^2}{g_\rho^2} 
%        \frac{g_\rho^2}{m_\omega^2 - i\Gamma_\omega m_\omega -s}
%        \left[
%           \frac{1}{-s} + \frac{3 g_{\rho\omega}}{e^2 g_\rho g_\omega}
%        \right],
        \frac{g_\omega}{g_\rho} 
        \frac{g_\rho^2}{m_\omega^2 - i\Gamma_\omega m_\omega -s}
        \left[
           \frac{g_\omega}{3} \frac{e^2}{-s} g_\rho + g_{\rho\omega}
        \right],
\label{eq:lab0}
\end{equation}
where $s$ denotes the invariant mass square of $\pi^+\pi^-$.
The size of the mass mixing ($g_{\rho\omega}$) can be
evaluated from
%\begin{equation}
%\left|
%  \frac{\vev{\pi^+ \pi^-|J_0^\mu|0}\epsilon_\mu}
%       {\vev{\pi^+ \pi^-|J_3^\mu|0}\epsilon_\mu}
%\right|^2
%  = 9 \frac{\Gamma(\omega\rightarrow e^+ e^-)}
%           {\Gamma(\rho\rightarrow e^+ e^-)}
%      \frac{\Gamma_\rho^2}{\Gamma_\omega^2}
%      \frac{\Gamma(\omega\rightarrow \pi^+\pi^-)}
%           {\Gamma(\rho\rightarrow \pi^+\pi^-)},
%\label{eq:lab4}
%\end{equation}
%where we have used an approximation $m_\rho\simeq m_\omega$.
\begin{equation}
\frac{\Gamma(\omega\rightarrow \pi^+ \pi^-)}
     {\Gamma(\rho  \rightarrow \pi^+ \pi^-)}
 = \frac{1}{(m_\rho^2-m_\omega^2)^2+\Gamma_\rho^2 m_\rho^2}
        \left[
           \frac{g_\omega}{3} \frac{e^2}{-m_\omega^2} g_\rho 
          + g_{\rho\omega}
        \right]^2.
\label{eq:lab4}
\end{equation}
Putting the experimental values, 
%$\Gamma (\omega\rightarrow e^+e^-)=0.6$keV, 
%$\Gamma (\rho\rightarrow e^+e^-)=6.8$keV,
$\Gamma (\omega\rightarrow \pi^+\pi^-)=0.19$MeV,
$\Gamma (\rho\rightarrow \pi^+\pi^-)=\Gamma_{\rho}=150$MeV,
and 
$\Gamma_{\omega}=8.4$MeV 
into Eq.(\ref{eq:lab4}), we find
%\begin{equation}
%  \frac{1}{3}\left[
%    \frac{1}{-m_\omega^2}+\frac{3g_{\rho\omega}}{e^2 g_\rho g_\omega}
%  \right] \simeq 0.57 \frac{\Gamma_\omega m_\omega}{e^2 g_\omega^2}.
%\label{eq:lab5}
%\end{equation}
\begin{equation}
  \frac{g_\omega}{3} \frac{e^2}{-m_\omega^2} g_\rho + g_{\rho\omega}
  \simeq 0.63 \Gamma_\omega m_\omega,
\label{eq:lab5}
\end{equation}
where the sign is determined from $e^+ e^- \rightarrow \pi^+ \pi^-$ 
near $\rho$-meson mass.

Plugging Eqs.(\ref{eq:lab0}) and (\ref{eq:lab5}) into
Eq.(\ref{eq:stphase}), we can evaluate the behavior of the strong
phase at the $\omega$ meson mass region,
%\begin{equation}
%  r e^{i\delta} \simeq 
%    r' e^{i\delta_q}
%    \frac{\beta e^{i\delta_\beta} 
%           (m_\omega^2 -i\Gamma_\omega m_\omega -s) 
%            +0.57\Gamma_\omega m_\omega}
%         {m_\omega^2 -i\Gamma_\omega m_\omega -s 
%            +0.57\alpha e^{i\delta_\alpha} \Gamma_\omega m_\omega},
%\end{equation}
%where we used $g_\rho \simeq g_\omega$.
\begin{equation}
  r e^{i\delta} \simeq 
    r' e^{i\delta_q}
    \frac{\beta e^{i\delta_\beta} 
           (m_\omega^2 -i\Gamma_\omega m_\omega -s) 
            +0.63\Gamma_\omega m_\omega}
         {m_\omega^2 -i\Gamma_\omega m_\omega -s 
            +0.63\alpha e^{i\delta_\alpha} \Gamma_\omega m_\omega}.
\end{equation}
In the ideal case, where $(\alpha, \beta, \delta_q)=(0,0,0)$, 
we thus obtain the maximum strong-phase difference at $s=m_\omega$.
% We thus obtain:
%\begin{eqnarray}
%  r 
%    &\simeq&r_p \sqrt{\frac{1+\alpha^2}{1+\beta^2}}
%        \sqrt{\frac{ [\beta(m_\omega^2-s)+0.57\Gamma_\omega m_\omega]^2
%                    +\beta^2 \Gamma_\omega^2 m_\omega^2}
%                   { [m_\omega^2 - s +0.57\alpha\Gamma_\omega m_\omega]^2
%                    +\Gamma_\omega^2 m_\omega^2}},
%    \\
%  \delta 
%    &\simeq& \tan^{-1}\frac{\Gamma_\omega m_\omega}
%                      {m_\omega^2 - s +0.57\alpha\Gamma_\omega m_\omega}
%       -\tan^{-1}\frac{\beta \Gamma_\omega m_\omega}
%                      {\beta(m_\omega^2 - s) +0.57\Gamma_\omega m_\omega}.
%\end{eqnarray}

We next roughly evaluate the sizes of these parameters $(\alpha, \beta,
r')$ of several decay modes.
The effective weak Hamiltonian is given by:
\begin{equation}
{\cal H}^T 
  = \frac{4G_F}{\sqrt{2}} V_{ub} V^*_{uq} 
      \sum_{i=1}^2 c_i(\mu) O_i^{(u)}, 
\qquad 
{\cal H}^P
  =-\frac{4G_F}{\sqrt{2}} V_{tb} V^*_{tq} 
      \sum_{i=3}^6 c_i(\mu) O_i,
\end{equation}
with
\begin{equation}
\begin{minipage}{0.45\textwidth}
\begin{eqnarray*}
O^{(u)}_1 &=& \bar q_{L\alpha} \gamma^\mu u_{L\beta}
              \bar u_{L\beta}  \gamma_\mu b_{L\alpha},
  \\
O_3       &=& \bar q_{L\alpha} \gamma^\mu b_{L\alpha}
              \sum \bar q'_{L\beta} \gamma_\mu q'_{L\beta},
  \\
O_5       &=& \bar q _{L\alpha} \gamma^\mu b_{L\alpha}
              \sum \bar q'_{R\beta} \gamma_\mu q'_{R\beta},
\end{eqnarray*}
\end{minipage}
\begin{minipage}{0.45\textwidth}
\begin{eqnarray*}
O^{(u)}_2 &=& \bar q_{L\alpha} \gamma^\mu u_{L\alpha}
              \bar u_{L\beta}  \gamma_\mu b_{L\beta},
  \\
O_4       &=& \bar q_{L\alpha} \gamma^\mu b_{L\beta}
              \sum \bar q'_{L\beta} \gamma_\mu q'_{L\alpha},
  \\
O_5       &=& \bar q _{L\alpha} \gamma^\mu b_{L\beta}
              \sum \bar q'_{R\beta} \gamma_\mu q'_{R\alpha},
\end{eqnarray*}
\end{minipage}
\end{equation}
with $\alpha$ and $\beta$ being color indices.
The Wilson coefficients of these operators at the $B$ meson mass
scale ($\mu=m_b=5$GeV) are calculated as\cite{kn:Desh,kn:Buras}:
\begin{equation}
\begin{minipage}{0.45\textwidth}
\begin{eqnarray*}
c_1(\mu) &=& -0.3125,
  \\
c_3(\mu) &=& 0.0174,
  \\
c_5(\mu) &=& 0.0104,
\end{eqnarray*}
\end{minipage}
\begin{minipage}{0.45\textwidth}
\begin{eqnarray*}
c_2(\mu) &=& 1.1502,
  \\
c_4(\mu) &=& -0.0373,
  \\
c_6(\mu) &=& -0.0459.
\end{eqnarray*}
\end{minipage}
\end{equation}
These coefficients receive finite renormalizations.
The finite radiative correction coming from ${\cal H}^T$ 
gives the following effects to the penguin-type on-shell quark
amplitude\cite{kn:Desh,kn:fleischer,kn:kramer}: 
\begin{equation}
  \sum_{i=3}^6 \vev{c_i(\mu) O_i}
  = \sum_{i=3}^6 c^{\rm eff}_i \vev{O_i}^{\rm tree},
\end{equation}
with
\begin{displaymath}
  c_3^{\rm eff} = c_3(\mu) -\frac{1}{3} P_s(k^2), \quad
  c_4^{\rm eff} = c_4(\mu) + P_s(k^2), \quad
  c_5^{\rm eff} = c_5(\mu) -\frac{1}{3} P_s(k^2), \quad
  c_6^{\rm eff} = c_6(\mu) + P_s(k^2),
\end{displaymath}
where $P_s$ is given by:
\begin{equation}
  P_s(k^2)
   = \frac{\alpha_s}{8\pi} c_2 \left[ 
        \frac{10}{9} + 4 \int_0^1 dx x(1-x) 
                         \ln\frac{m_c^2 - x(1-x)k^2}{\mu^2}
      \right],
\label{eq:ps}
\end{equation}
with $m_c$ and $k^2$ being charm quark mass and the gluon momentum,
respectively.  
We neglect any finite radiative corrections coming from penguin
Hamiltonian, as well as finite corrections to the tree amplitude, since they
are suppressed by $\alpha_s/4\pi$.
The choice of $k^2$ is important for evaluating the strong
phase at the quark level.
In this work, we use $k^2/m_b^2=0.5$ and $k^2/m_b^2=0.3$.

%In this work, we take $k^2=m_b^2/2$ and 
%neglect $m_q^2/k^2$ term in Eq.(\ref{eq:ps}).
 
%These finite renormalizations $m_{ij}$ contain imaginary part
%(strong phase, final state interaction) comming from the quark loop
%diagram.
%%We neglect the effect of $m_{ij}$ in the following calculation.
%Although the imaginary part in $m_{ij}$ can trigger the $B^{\pm}$
%decay $CP$ asymmetry, we neglect this effect for a while and
%concentrate on the $CP$ asymmetry caused by $\rho$-$\omega$
%interference. 
%We thus neglect the effect of $m_{ij}$ in the following calculation.

We neglect the effects of the electroweak penguin operators in this paper,
since they are smaller than the uncertainties of the hadronic
matrix estimation.

We start from the $B^-\rightarrow \rho^0 \rho^-$ decay.
Within the factorization approximation of the hadronic matrix elements, 
we obtain:
\begin{eqnarray*}
\vev{\rho^0 \rho^- | \sum_{i=1}^2 c_i O_i | B^-}
  &=& (c_1 + c_2/N_c) \vev{\rho^0| \bar u_L\gamma^\mu u_L | 0}
                 \vev{\rho^-| \bar d_L\gamma_\mu b_L | B^-}
  \nonumber\\
  & &
     +(c_1/N_c + c_2) \vev{\rho^-| \bar d_L\gamma^\mu u_L | 0}
                 \vev{\rho^0| \bar u_L\gamma_\mu b_L | B^-},
  \\
\vev{\omega \rho^- | \sum_{i=1}^2 c_i O_i | B^-}
  &=& (c_1 + c_2/N_c) \vev{\omega| \bar u_L\gamma^\mu u_L | 0}
                 \vev{\rho^-| \bar d_L\gamma_\mu b_L | B^-}
  \nonumber\\
  & &
     +(c_1/N_c + c_2) \vev{\rho^-| \bar d_L\gamma^\mu u_L | 0}
                 \vev{\omega| \bar u_L\gamma_\mu b_L | B^-},
\end{eqnarray*}
where we have neglected the annihilation diagram.
Since the factorization approximation is not theoretically 
justified, we leave the color suppression factor ($1/N_c$) as a 
free parameter.
Actually, measurements of the branching fractions of $B\rightarrow D$
decays indicate $(c_1+c_2/N_c)/(c_2+c_1/N_c)\simeq 0.15$, which
implies $N_c\simeq 2$ in the factorization assumption\cite{kn:Rod96}.
In this work, we use $N_c=2$ and $N_c=\infty$.

It is convenient to define
\begin{displaymath}
  A(\epsilon_c, q_c, \epsilon_n, q_n)
  \equiv \vev{\rho^0|\bar u_L \gamma^\mu u_L| 0}
         \vev{\rho^-|\bar d_L \gamma_\mu b_L| B^-},
\end{displaymath}
where $\epsilon_c$ and $\epsilon_n$ are the polarization vector of
$\rho^-$ and $\rho^0$, respectively, and $q_c$ and $q_n$ are the
four momentum of the $\rho^-$ and $\rho^0$, respectively.
By using the Lorentz structure of the $\rho$-meson form factor\cite{bsw}
$\vev{\rho^-|\bar d_L \gamma_\mu b_L| B^-}$, we can show
$
  A(\epsilon_c, q_c, \epsilon_n, q_n)
  \equiv 
  A(\epsilon_n, q_n, \epsilon_c, q_c)
$.
The isospin $U(2)$ symmetry thus leads to
\begin{eqnarray*}
\vev{\rho^0 \rho^- | \sum_{i=1}^2 c_i O_i | B^-}
  &=& (1+1/N_c) (c_1 + c_2) A(\epsilon_c, q_c, \epsilon_n, q_n),
  \\
\vev{\omega \rho^- | \sum_{i=1}^2 c_i O_i | B^-}
  &=& (1+1/N_c) (c_1 + c_2) A(\epsilon_c, q_c, \epsilon_n, q_n).
\end{eqnarray*}

The contributions from the penguin Hamiltonian can be evaluated in a
similar manner:
\begin{eqnarray}
\vev{\rho^0 \rho^- | \sum_{i=3}^6 c_i O_i | B^-}
  &=& 0
  \\
\vev{\omega \rho^- | \sum_{i=3}^6 c_i O_i | B^-}
  &=& 2((1+1/N_c) c_3 + (1+1/N_c) c_4 + c_5 + c_6/N_c) 
      A(\epsilon_c, q_c, \epsilon_n, q_n).
\end{eqnarray}
We thus obtain
\begin{equation}
  \alpha e^{i\delta_\alpha} =1, \quad 
  \beta e^{i\delta_\beta} =0,  \quad
  r'e^{i\delta_q} = 2
         \frac{(1+1/N_c) c_3 + (1+1/N_c) c_4 + c_5 + c_6/N_c}
                {(1+1/N_c) (c_1 + c_2)}
         \left|\frac{V_{tb} V^*_{td}}{V_{ub} V^*_{ud}}\right|.
\end{equation}
We note that the $\rho$-meson polarization dependences
on the hadronic parameters, e.g., $\alpha$, $\beta$,
disappear in this $U(2)$ symmetry
approximation. 
The $CP$-odd angular coefficients\cite{kn:kramer} do not appear within
this assumption.

In a similar manner, we evaluate the parameters of
the $B^-\rightarrow \rho^0 K^{*-}$ decay:
\begin{eqnarray}
  & &
  \alpha e^{i\delta_\alpha} = 1, \quad 
  \beta e^{i\delta_\beta} =\frac{c_3/N_c + c_4}
             {(2+1/N_c) c_3 + (1+2/N_c) c_4 + 2c_5 + 2c_6/N_c},
  \nonumber\\
  & &
  r'e^{i\delta_q} = 
         \frac{(2+1/N_c) c_3 + (1+2/N_c) c_4 + 2c_5 + 2c_6/N_c}
                {(1+1/N_c) (c_1 + c_2)}
         \left|\frac{V_{tb} V^*_{ts}}{V_{ub} V^*_{us}}\right|,
\end{eqnarray}
where we have neglected the $U(3)$ breaking effects in the form factors.

For the $B^- \rightarrow \rho^0 \pi^-$ and $B^- \rightarrow \rho^0
K^-$ decays, 
we evaluate the matrix elements by using the form factors calculated in
Ref.\cite{bsw}.

The numerical results are summarized in Tables 
\ref{table1}, \ref{table2}, \ref{table3}, and \ref{table4} for 
$N_c=2,\infty$ and $k^2/m_b^2=0.5,0.3$.
We used $m_u=5$MeV, $m_d=8$MeV, $m_s=150$MeV, $m_c=1.35$GeV and
$m_b=5$GeV in this calculation.

We next discuss how we can extract the weak phase from
the observed $CP$ asymmetries.
The existence of the short distance absorptive part triggers
a $CP$ asymmetry in $\Gamma(B^\pm \rightarrow \rho^0)$ and
$\Gamma(B^\pm \rightarrow \omega)$:\cite{kn:kramer}
\begin{eqnarray}
\frac{\Gamma(B^- \rightarrow \rho^0) - \Gamma(B^+ \rightarrow \rho^0)}
     {\Gamma(B^- \rightarrow \rho^0) + \Gamma(B^+ \rightarrow \rho^0)}
&=& \frac{\beta r' ( \cos(\delta_q + \delta_\beta + \phi)
                    -\cos(\delta_q + \delta_\beta - \phi) ) }
         {1+ \beta^2 r^{\prime 2} 
          +\beta r'( \cos(\delta_q + \delta_\beta + \phi)
                    -\cos(\delta_q + \delta_\beta - \phi) ) },
\label{eq:rhoasym}
\\
\frac{\Gamma(B^- \rightarrow \omega) - \Gamma(B^+ \rightarrow \omega)}
     {\Gamma(B^- \rightarrow \omega) + \Gamma(B^+ \rightarrow \omega)}
&=& \frac{\alpha^{-1} r' ( \cos(\delta_q - \delta_\alpha + \phi)
                    +\cos(\delta_q - \delta_\alpha - \phi) ) }
         {1+ \alpha^{-2} r^{\prime 2} 
          +\alpha^{-1} r'( \cos(\delta_q - \delta_\alpha + \phi)
                    +\cos(\delta_q - \delta_\alpha - \phi) ) }.
\label{eq:omegasym}
\end{eqnarray}
We also obtain the ratio of partial decay width:
\begin{equation}
\frac{\Gamma(B^- \rightarrow \omega) + \Gamma(B^+ \rightarrow \omega)}
     {\Gamma(B^- \rightarrow \rho^0) + \Gamma(B^+ \rightarrow \rho^0)}
 = \frac{\alpha^2+ r^{\prime 2} 
          +\alpha r'( \cos(\delta_q - \delta_\alpha + \phi)
                    +\cos(\delta_q - \delta_\alpha - \phi) ) }
        {1+ \beta^2 r^{\prime 2} 
          +\beta r'( \cos(\delta_q + \delta_\beta + \phi)
                    +\cos(\delta_q + \delta_\beta - \phi) ) }.
\label{eq:rate}
\end{equation}

%Let us consider $\rho^-\rho^0(\omega)$ mode, in which the hadronic
%parameters are simplified by the isospin $U(2)$ symmetry: $\alpha=1$, 
%$\delta_\alpha=0$ and $\beta=0$.
%Thus, we have two hadronic parameters $r'$, $\delta_q$ and one weak
%phase $\phi_2={\rm arg}((V_{ud}V^*_{ub})/(V_{td}V^*_{tb}))$ in this
%decay mode. 
%Since the $\rho$ meson asymmetry Eq.(\ref{eq:rhoasym}) vanishes 
%in this mode, the above mentioned observables Eq.(\ref{eq:omegasym})
%and Eq.(\ref{eq:rate}) are not enough to determine whole parameters
%$r'$, $\delta_q$ and $\phi_2$. 
%If we can measure the $CP$ asymmetry near the $\omega$ meson pole
%Eq.(\ref{eq:cpasym}) precisely, we have sufficient
%informations to extract the weak phase $\phi_2$.
%Although the existence of the electroweak penguin operators causes
%$\beta\ne0$ and makes the problem complicated, we can estimate the
%size of them from the $CP$ asymmetry of $B^{\pm}\rightarrow
%\rho^{\pm}\rho^0$.  

Since the quark loop absorptive parts of the tree amplitudes are
negligibly small, we can assume that $\delta_\alpha=0$.
We therefore have six unknown parameters
$(\alpha,~\beta,~\delta_{\beta},~\delta_q,~r',~\phi_i)$.
On the other hand, in addition to the above-mentioned observables,
Eqs.(\ref{eq:rhoasym})--(\ref{eq:rate}), 
we can measure two $M(\pi^+\pi^-)$ spectra ( for $B^-$ and $B^+$ )
around $M(\omega)\pm\Gamma(\omega)$ due to $\rho-\omega$ interference;
each provides
two types of information, i.e., the pole position and magnitude.
We can, therefore, derive the weak phase $\phi_i~(i=2,3)$ from these
measurements without assuming any theoretical models of the hadron
matrix elements. 
We also note that the existence of the electroweak penguin operators
does not affect this determination.
For most of decay modes we can also assume $\alpha=1$, which
improves the accuracy of the $\phi_i$ determination.

We demonstrate the asymmetry patterns ($\pi^+\pi^-$ invariant
mass spectra) in Figures \ref{pattern} (a1)-(d3),
%\begin{figure}
%\epsfbox{pattern.eps}
%\caption{Expected invariant mass spectra of unlike-sign pion pairs.
%The solid lines are for $B^+$ decays and the dashed ones are for $B^-$
%decays. The vertical scale is the expected yield for a 
%$(\pi^+\pi^-)h^{\pm}$
%combination. Various $\alpha$s and $\beta$s are assumed; (a)
%($\alpha$, $\beta$)=(1, 0), (b) (1, 0.3), (c) (1.3, 0), 
%and (d) (1.3, 1.3).}
%\label{pattern}
%\end{figure}
where (a), (b), (c), and (d) denote $B^-\rightarrow \rho^0\rho^+$,
$\rho^0 K^{*-}$, $\rho^0 K^-$, and $B^0\rightarrow \rho^0 \bar{K}^0$
decay modes,
respectively.
The results using the numerical values of Table \ref{table1} are given in
Figures \ref{pattern} (a1), (b1), (c1), and (d1) and
those of Table \ref{table3} are shown in Figures \ref{pattern} (a3), (b3),
(c3), and (d3).
In Figures \ref{pattern} (a2), (b2), (c2), and (d2), we used ``zero
hadronic phases" (i.e., 
$\delta_{\beta},\delta_q=$0 or $\pm\pi$) with $\alpha$, $\beta$, and
$r'$ from Table \ref{table1}.
The solid lines are for $B^+$ or $B$ and the dashed ones are
for $B^-$ or $\bar{B}$. 
Here, we have assumed the KM matrix of Wolfenstein parametrization,
$\lambda=0.221$, $\rho=-0.12$, $\eta=0.34$ and $A=0.84$, which
corresponds to  
%sin($\theta_C$)=0.221,
%$\sqrt{\rho^2+\eta^2}=0.36$, 
$(\phi_1,\phi_2,\phi_3)=(15,55,110)$ degrees.
The branching ratios in these parameters are given in Table \ref{table5}.
The vertical scales are normalized to give the
number of entries at $10^8$ $B\bar{B}$ events with an 100-\% acceptance.
Drastic asymmetries appear around the $\omega$ mass region.

We list the asymmetries obtained for the various decay modes in Table
\ref{table5}. The hadronic parameters in Tables \ref{table1} --
\ref{table4} 
are assumed. $A(\rho^0)$ and $A(\omega)$ are the asymmetries
of the $\rho^0 h$ and $\omega  h$ modes.
$A(\rho \omega)$ is the mean asymmetry of the $M(\pi^+\pi^-)$
invariant mass spectra around $M(\omega)\pm\Gamma(\omega)$,
and $A^{max}(\rho \omega)$ is the maximum asymmetry in this
region. $A^0(\rho \omega)$ is obtained by assuming the
``zero hadronic phase".
The branching ratios are also estimated using this formalism.
The $CP$ asymmetries via $\rho$-$\omega$ interference are large
($>$ 10\%) in most of cases.

In order to check the feasibility for detecting this $CP$
asymmetry,
we performed a simulation assuming the BELLE detector of 
the KEK B-factory\cite{belle}, an asymmetric $e^+e^-$ collider
(8 x 3.5GeV). 
The invariant mass 
resolution of $\pi^+\pi^-$ around $\omega$ mass is expected to be
3.2 MeV for the $B\rightarrow \omega h$, $\omega\rightarrow \pi^+\pi^-$
decay; this is enough to resolve the interference pattern.
Here, the momentum resolution is derived from
$(dP_T/P_T)^2=(0.001P_T/1{\rm GeV})^2+0.002^2$. 
In the case of a symmetric collider, the mass resolution will be
better.
In case of hadron machine, the average $B$ mesons' $P_T$ would be
several GeV or more. 
Although the mass resolution slightly deteriorates,
the statistics are sufficient in hadron machines.

%Here, we carried out a simple analysis assuming a typical detector at
%an asymmetric $e^+e^-$ collider such as KEK-B-factory \cite{belle}.
In order to suppress
the large background from continuum events under $\Upsilon (4S)$,
we used two cuts in analysing the $\rho^0 h^{\pm,0}$ decay
\cite{cleo}: one was that the 
absolute value of the cosine
of the angle between the thrust axes of B decay products 
and the other particles
at center-of-mass-system of $\Upsilon (4S)$ be less than $0.6$;
the other was that the energy of the $B$ candidate be between 
$5.25$ and $5.325$ GeV. 
The combined efficiency of these two cuts was 60\%.
The beam-energy constraint mass spectra were used.
The acceptances of $\rho^0\pi^{\pm}$ and $\rho^0K^{\pm}$ were 
found to be $\sim$35\%.
The efficiency of charged track is typically 80\% and that of $\pi^0$
$\sim$33\%.
The backgrounds from $\Upsilon (4S)$ decay, i.e., $B\bar{B}$ events,
are mostly $h^-\bar{D}(\bar{D}\rightarrow K^-\pi^+,K^-\rho^+,...)$.
We can, therefore, reject combinations if there are possible 
combination including $K^-$, which have masses consistent with
$D$ or $D^*$. This cut does not deteriorate the acceptances.

%We tried to simulate such decay modes as $B\rightarrow \rho^0h^-$ 
%($h^-=\pi^-,\rho^-,K^-,K^{*-}$).
The results of a simulation for $B^{\pm}\rightarrow \rho^0h^{\pm}$
($h^{\pm}=\pi^\pm,\rho^\pm,K^\pm,K^{*\pm}$) decay modes are 
summarized in $N(B\bar B)$ of Table \ref{table5}, 
the necessary number of $B\bar B$ events for detecting the 3$\sigma$ $CP$
asymmetry at the $\rho$-$\omega$ interference region. 
The branching ratios quoted in Table \ref{table5} are assumed.
%Necessary numbers of $B\bar{B}$ events ($N(B\bar{B})$) are listed
%in Table \ref{table5}.
In some of these decay modes, 3$\sigma$-CP violations
are detectable with $10^9$ $B\bar{B}$ events
by $\rho-\omega$ interference modes.
A luminosity of $10^{35}cm^{-2}s^{-1}$ is necessary.
It may be possible if the experimental setup (including
accelerator's setup) is optimized for the detection of
direct $CP$ violations.
Also, in these decay modes, direct $CP$
violations via $B\rightarrow \omega h,\omega\rightarrow \pi^+\pi^-\pi^0$
can be detected with the same luminosity
if their asymmetries are on the order of
10\% or larger.
This is because both the $\rho$-$\omega$ interference and the $\omega$-peak
are narrow, which reduces the background.
If a good method can be found to suppress the background from the continuum,
the necessary luminosity can be significantly reduced.
This is due to large $CP$ asymmetries via $\rho-\omega$ interference,
which is less sensitive to theoretical assumptions.
If these things are realized, we can determine $\phi_i~(i=2,3)$,
as we previously discussed.

%discussion
The advantage of this method are as follows:
\begin{itemize}
\item 
The behavior of the final state interaction can be controlled by using
the $\omega$ meson pole. 
This method is insensitive to the ambiguity of other hadronic phases,
which can be calculated from experimental data.  
%We only looked at the $\pi^+\pi^-$ invariant mass around $\omega$ mass.
\item 
$\phi_2$ and $\phi_3$ can be measured by the Cabibbo-allowed and
Cabibbo-suppressed channels, respectively. 
We also note that the direct $CP$ asymmetry discussed in this paper is
proportional to $\sin \phi_i$, which makes it possible to measure the
weak phases for $\phi_i \simeq \pi/2$.
In the case of indirect $CP$ violation, asymmetry is proportional to
$\sin 2\phi_i$ \cite{sanda}.
% In this case asymmetry is proportional to $\sin \phi_i$
%($i$=2,3) and presently they are estimated to be around 90 degrees.
\item
The typical experimental detector has some material in front of the central
tracking device. This causes a systematic difference between
the $h^+$ and $h^-$ yields. 
The present method looks only at the $M(\pi^+\pi^-)$
shape, i.e., insensitive to these systematic effects.
Only statistics and a high mass resolution are necessary.
\item
No tagging of the other B meson, nor the vertex chambers, to measure
such a small decay length as $\sim$pico-second of $B$ meson is essential.
\end{itemize}

%conclusion
We have studied the effect of $\rho$-$\omega$ interference in the
decay modes $B\rightarrow \rho^0 (\omega) h$, 
$\rho^0 (\omega) \rightarrow \pi^+\pi^-$,
where $h$ is any hadronic final state, such as
$\pi$, $\rho$, or $K$.
%$\omega$ meson decays to $\pi^+\pi^-$ by BR=2.2\%
%which is isospin violating. 
Although the isospin-violating decay of $\omega\rightarrow\pi^+\pi^-$ 
is a small effect with BR=$2.2$\%, 
the interference at the kinematical region $M(\pi^+\pi^-)\sim
M(\omega)\pm \Gamma_{\omega}$ is enhanced by the $\omega$ pole.
%when we restrict ourselves a 
%kinematical
%region to $M(\pi^+\pi^-)\sim M(\omega)\pm \Gamma_{\omega}$,
%the interference is enhanced by a pole-dominance effect.
We have shown the $CP$ asymmetry to be 
sufficiently large to be detected.
The $CP$ asymmetry appears in the deformation of the
Breit-Wigner shape of the $\rho^0\rightarrow \pi^+\pi^-$
invariant mass spectrum. 
The prediction of the $CP$ asymmetry is not very sensitive to the
hadronic phase calculation, i.e., a ``sure" prediction.
Any B-factory, even if it is a symmetric $e^+e^-$ collider
or hadron machine, can carry out this measurement.
We only need to accumulate enough statistics
and to have a mass
resolution [$\Delta M(\pi^+\pi^-)$] better than
the width of the $\omega$ meson (8.4MeV) at around the $\omega$ mass region.
An attempt to obtain more than $10^9$ $B\bar{B}$ is important.

%acknowledgement
We thank Drs. M. Kobayashi, A. I. Sanda, M. Tanaka and I. Dunietz
for useful discussions. 
We also thank Belle collaboration for providing detector simulation
programs.
M.T. is supported from Japanese Ministry of Education, Science and
Culture for his stay at Fermilab. 
He thanks Fermilab Theoretical Physics Group for hospitality.

%\end{document}
\begin{table}
\begin{tabular}{l|lllll}
  & $\alpha$ & $\beta$ & $\delta_\beta$ & $r'$ & $\delta_q$ \\
\hline
$B^- \rightarrow \rho^0 \rho^-$ 
  & 1. & 0.       &  ---   &  $9.6\times 10^{-2} \vd$  & $-2.807$
  \\
$B^- \rightarrow \rho^0 K^{*-}$ 
  & 1. & 0.524    &  0.020 &  $6.3\times 10^{-2} \vs$  & $-2.814$
  \\
$B^- \rightarrow \rho^0 \pi^-$ 
  & 1. & $0.423$ &  3.103 &  $1.0\times 10^{-1} \vd$  & $-2.796$
  \\
$B^- \rightarrow \rho^0 K^-$ 
  & 1. & $0.117$  &  0.343 &  $4.8\times 10^{-2} \vs$  & $-2.792$
  \\
$\bar B^0 \rightarrow \rho^0 \pi^0$
  &0.262 & $2.035$& $3.138$ & $9.5\times 10^{-2} \vd$  & $-2.805$
  \\
$\bar B^0 \rightarrow \rho^0 \bar K^0$
  &1.  & $0.126$ & $-2.830$ & $1.63\times 10^{-1}\vs$  & $-2.793$
\end{tabular}
\caption{$N_c=2$, $k^2=0.5 m_b^2$}
\label{table1}
\end{table}
\begin{table}
\begin{tabular}{l|lllll}
  & $\alpha$ & $\beta$ & $\delta_\beta$ & $r'$ & $\delta_q$ \\
\hline
$B^- \rightarrow \rho^0 \rho^-$ 
  & 1. & 0.       &  ---  &  $3.5\times 10^{-2} \vd$  & $-2.749$
  \\
$B^- \rightarrow \rho^0 K^{*-}$ 
  & 1. & 2.270    &$-3.060$&  $2.8\times 10^{-2} \vs$  & $ 0.247$
  \\
$B^- \rightarrow \rho^0 \pi^-$ 
  & 1. & 3.082    &$0.183$ &  $3.9\times 10^{-1} \vd$  & $ 0.122$
  \\
$B^- \rightarrow \rho^0 K^-$ 
  & 1. & 0.126    &$-2.964$&  $1.64\times 10^{-1}\vs$  & $ 0.284$
  \\
$\bar B^0 \rightarrow \rho^0 \pi^0$
  & 0.262 &  1.413&$0.030$ &  $1.44\times 10^{-1}\vd$  & $-2.851$
  \\
$\bar B^0 \rightarrow \rho^0 \bar K^0$
  & 1.    & 0.136 & $0.168$&  $2.13\times 10^{-1}\vs$  & $-2.858$
\end{tabular}
\caption{$N_c=\infty$, $k^2=0.5 m_b^2$}
\label{table2}
\end{table}
\begin{table}
\begin{tabular}{l|lllll}
  & $\alpha$ & $\beta$ & $\delta_\beta$ & $r'$ & $\delta_q$ \\
\hline
$B^- \rightarrow \rho^0 \rho^-$ 
  & 1. & 0.       &  ---   &  $1.1\times 10^{-1} \vd$  & $-3.056$
  \\
$B^- \rightarrow \rho^0 K^{*-}$ 
  & 1. & 0.527    & 0.005  &  $7.2\times 10^{-2} \vs$  & $-3.058$
  \\
$B^- \rightarrow \rho^0 \pi^-$ 
  & 1. & 0.420    & 3.133  &  $1.15\times 10^{-1}\vd$  & $-3.054$
  \\
$B^- \rightarrow \rho^0 K^-$ 
  & 1. & 0.119    & 0.075  &  $5.5\times 10^{-2} \vs$  & $-3.053$
  \\
$\bar B^0 \rightarrow \rho^0 \pi^0$
  & 0.262 &  2.033 &  3.141&  $1.1\times 10^{-1} \vd$  & $-3.056$
  \\
$\bar B^0 \rightarrow \rho^0 \bar K^0$
  & 1. & 0.128   & $-3.074$&  $1.87\times 10^{-1}\vs$  & $-3.053$
\end{tabular}
\caption{$N_c=2$, $k^2=0.3 m_b^2$}
\label{table3}
\end{table}
\begin{table}
\begin{tabular}{l|lllll}
  & $\alpha$ & $\beta$ & $\delta_\beta$ & $r'$ & $\delta_q$ \\
\hline
$B^- \rightarrow \rho^0 \rho^-$ 
  & 1. & 0.       & ---      &  $4.1\times 10^{-2} \vd$ & $-3.043$
  \\
$B^- \rightarrow \rho^0 K^{*-}$ 
  & 1. & 2.323    & $-3.123$ & $3.1 \times 10^{-2} \vs$ & $0.065$
  \\
$B^- \rightarrow \rho^0 \pi^-$ 
  & 1. & $3.291$  & $0.045$  & $4.2\times 10^{-2} \vd$  & $0.034$
  \\
$B^- \rightarrow \rho^0 K^-$ 
  & 1. & $0.131$  & $-3.102$ & $1.86\times 10^{-1} \vs$ & $0.074$
  \\
$\bar B^0 \rightarrow \rho^0 \pi^0$
  & 0.262 & 1.424 & $0.007$  & $1.63\times 10^{-1}\vd$  & $-3.066$
  \\
$\bar B^0 \rightarrow \rho^0 \bar K^0$
  &1.  & $0.140$  & $0.038$  & $2.43\times 10^{-1}\vs$  & $-3.068$
\end{tabular}
\caption{$N_c=\infty$, $k^2=0.3 m_b^2$}
\label{table4}
\end{table}
\begin{table}
\begin{tabular}{l|rrrrrrrr}
Mode & Table & BR & $A(\rho^0)$ & $A(\omega)$ & $A(\rho \omega)$
& $A^{max}(\rho \omega)$ & $A^0(\rho \omega)$ & $N(B\bar{B})$\\
 & & $\times 10^{-8}$ & \% & \% & \% & \% & \% & $\times10^8$\\
\hline
$B^- \rightarrow \rho^0 \rho^-$ & I&
2100 & 0 & 10 & 13 & 26 & 11 & 70\\
$B^- \rightarrow \rho^0 K^{*-}$ & I&
720 & -36 & -19 & -45 & -79 & -19 & 12\\
$B^- \rightarrow \rho^0 \pi^-$ & I&
660 & -6 & 11 & 16 & 37 & 18 & 29\\
$B^- \rightarrow \rho^0 K^-$ & I&
62 & -41 & -26 & -82 & -91 & -67 & 7.6\\
\hline
$B^- \rightarrow \rho^0 \rho^-$ & II &
910 & 0 & 5 & 5 & 9 & 4 & 1900\\
$B^- \rightarrow \rho^0 K^{*-}$ &II&
840 & -19 & 16 & 10 & 32 & 28 & 490\\
$B^- \rightarrow \rho^0 \pi^-$ &II&
170 & -21 & -2 & -5 & -17 & 13 &2600\\
$B^- \rightarrow \rho^0 K^-$ &II&
33 & -59 & 5 & 33 & 58 & 31 & 15\\
\hline
$B^- \rightarrow \rho^0 \rho^-$ &III&
2100 & 0 & 3 & 14 & 30 & 13 & 61\\
$B^- \rightarrow \rho^0 K^{*-}$ &III&
780 & -9 & -4 & -23 & -57 & -17 & 42 \\
$B^- \rightarrow \rho^0 \pi^-$ &III&
660 & -2 & 3 & 20 & 48 & 21 & 16\\
$B^- \rightarrow \rho^0 K^-$ &III&
56 & -13 & -6 & -70 & -82 & -65 & 11\\
\hline
$B^- \rightarrow \rho^0 \rho^-$ &IV&
910 & 0 & 2 & 5 & 11 & 5 & 1900\\
$B^- \rightarrow \rho^0 K^{*-}$ &IV&
970 & -4 & 4 & 20 & 40 & 24 & 98\\
$B^- \rightarrow \rho^0 \pi^-$ &IV&
170 & -31 & -7 & -12 & -27 & 16 & 440\\
$B^- \rightarrow \rho^0 K^-$ &IV&
33 & -15 & 1 & 28 & 48 & 28 & 17
\end{tabular}
\caption{
Asymmetries for the various decay modes;
Table denotes the Table number for hadronic parameters,
and BR is branching ratios in unit of $10^{-8}$.
$A(\rho^0)$ and $A(\omega)$ are asymmetries for
$B^-\rightarrow \rho^0 h$ and $\omega h$ modes, respectively.
$A(\rho \omega)$ is that in the region of $M(\omega)\pm
\Gamma(\omega)$. $A^{max}(\rho \omega)$ is the maximum
asymmetry in this region. $A^0(\rho \omega)$ is that under assumption
of ``zero hadronic phase".
$N(B\bar{B})$ is a necessary number of $B\bar{B}$ events
in order to obtain a 3$\sigma$ asymmetry in $\rho-\omega$ interference
region.
}
\label{table5}
\end{table}

\end{document}